\documentclass[preprint,12pt]{elsarticle}

\usepackage{graphicx}

\usepackage{mathptmx}
\usepackage{float}

\begin{document}

\begin{frontmatter}

\title{The Disordered Induced Interaction and the Phase Diagram of Cuprates}

\author{E. V. L. de Mello and Raphael B. Kasal}

\address{%
Instituto de F\'{\i}sica, Universidade Federal Fluminense, Niter\'oi, RJ
24210-340, Brazil\\}%

\begin{abstract}

There are processes in nature that resemble a true force but arise due to the minimization 
of the local energy. The most well-known case is the exchange interaction that leads to 
magnetic order in some materials. We discovered a new similar process occurring in 
connection with an electronic phase separation transition that leads to charge 
inhomogeneity in cuprate superconductors.  The minimization of the local free energy,
described here by the Cahn-Hilliard diffusion equation, 
drives the charges into regions of low and high densities. This motion leads to an 
effective potential with two-fold effect: creation of tiny isolated regions or 
micrograins, and two-body attraction, which promotes local or intra-grain 
superconducting pairing. Consequently, as in granular superconductors, the 
superconducting transition appears in two steps. First, with local intra-grain 
superconducting amplitudes  and, at lower temperature, the superconducting 
phase or resistivity transition is attained by intergrain Josephson coupling.  
We show here that this approach reproduces the main features of the cuprates phase diagram, gives 
a clear interpretation to the pseudogap phase and yields the position dependent 
local density of states gap $\Delta(\vec r)$ measured by tunnelling experiments.
\end{abstract}


\end{frontmatter}

\section{Introduction}

Almost 25 years of intense research
on copper-oxide-based high-temperature superconductors has
revealed many interesting and non-conventional results. It is
impossible to explained all these data in detail by a single
theory because some features may be sample dependent, but there
are some general properties believed to be common to all
cuprate superconductors like the pseudogap phase and the d-wave
superconducting amplitude symmetry\cite{TS,Tallon,Sawatzky}. On materials
which allow surface studies and thin films, mostly Bi, La and
Y-based cuprates, there are well measured properties: the larger gap
at the leading edge of the Fermi surface or antinodal (along the
$Cu-O$ bonds) direction (($\pm \pi,0$) and ($0,\pm
\pi$))\cite{Damascelli}, with the consequent nodal Fermi arcs above
the critical temperature $T_c$\cite{Lee,Chatterjee}, and the
spatially dependent local density of states gap $\Delta(\vec r)$
measured by atomically resolved spectroscopy such as scanning
tunneling microscopy (STM)\cite{McElroy,Gomes,Pasupathy,Kato,Pushp,Kato2}.

We show here that the fundamental superconducting interaction 
and some of these properties can be interpreted
evoking an electronic phase separation (EPS) and its consequent
temperature evolution of the charge inhomogeneity. There are many
evidences that the charge distribution in the $CuO_2$ planes of the
high temperature superconductors (HTSC) is microscopically
inhomogeneous\cite{Muller,Mello03}. Several different experiments like neutron
diffraction\cite{Tranquada,Bianconi,Bozin}, muon spin relaxation
($\mu SR$)\cite{Uemura,Sonier}, nuclear quadrupole resonance (NQR) 
and nuclear magnetic resonance NMR\cite{Singer,Keren,Curro}
have  detected varying local electronic densities. The spatial
variations of the  electronic gap amplitude $\Delta_{sc}(\vec r)$ at
a nanometer length scale\cite{McElroy,Gomes,Pasupathy,Kato,Pushp,Kato2} may also
be connected with the charge inhomogeneities.
The origin of this electronic disorder is still not clear; it may be
from the quenched disorder introduced by the dopant
atoms\cite{Keren,Bianconi2}, or it may be due to competing orders in
the $CuO_2$ planes\cite{Mello09}. Probably, the most convincing evidence of an EPS
transition came from the increase of the local doping difference with
decreasing temperature in the NQR experiment of Singer et al\cite{Singer}.
An EPS has been also used to interpret transport
properties\cite{Salluzzo,Felipe} and to provide a clear interpretation
to the non-vanishing  magnetic susceptibility  measured above the 
resistivity transition $T_c$\cite{Cabo,JL2007}. More recently
a phase separation in $La_2CuO_{4+y}$ was analyzed after different
times of the annealing processes\cite{Bianconi2}

The cuprates phase diagram has a crossover temperature or
upper pseudogap detected by several transport experiments that
merges with $T_c$ in the overdoped region\cite{TS,Tallon}, i.e.,
for large dopant average level $p$. It is our fundamental assumption, and
the starting point of our theory, that such crossover line is
related to the EPS transition temperature $T_{PS}(p)$. Phase
separation is a very general phenomenon in which a structurally and
chemically homogeneous system shows instability toward a disordered
composition\cite{Bray}. However, the problem is to describe and to
follow quantitatively the time and temperature evolution of the
charge separation process. We have shown that an appropriate way to
do this is through the general Cahn-Hilliard (CH)
theory\cite{CH,Otton,Mello04}.

This approach uses an order parameter that is the difference
between the temperature dependent local doping concentration
$p(i,T)$ and the average doping level $p$, i.e., $u(i,T)\equiv (p(i,T)-p)/p$.
Then the local Ginzburg-Landau (GL) free energy functional is,
\begin{eqnarray}
f(i,T)= {{{1\over2}\varepsilon^2 |\nabla u(i,T)|^2 +V_{GL}(u(i,T))}}.
\label{FE}
\end{eqnarray}
Where the potential ${\it V}_{GL}(u,i,T)= -A^2(T)u^2/2+B^2u^4/4+...$,
$A^2(T)=\alpha(T_{PS}(p)-T)$, $\alpha$ and $B=1$ are temperature
independent parameters. $\varepsilon$ gives
the size of the grain boundaries among two distinct
phases\cite{Otton,Mello04}.

The CH equation can be derived\cite{Bray}
from a continuity equation of the local free energy $f(i,T)$,
$\partial_tu=-{\bf \nabla.J}$, with the current ${\bf J}=M{\bf
\nabla}(\delta f/ \delta u)$, where $M$ is the mobility or the
charge transport coefficient, normally incorporated in the
time step intervals. Therefore,
\begin{eqnarray}
\frac{\partial u}{\partial t} = -M\nabla^2(\varepsilon^2\nabla^2u
+ {\it V}_{GL}(p,i,T)).
\label{CH}
\end{eqnarray}
We have already made a detailed study of the  CH differential equation by
finite difference methods\cite{Otton} which yields the density profile
$u(p,i,T)$ in a $105\times 105$ array as function of the time steps,
up to the stabilization of the local
densities\cite{Mello04,DDias07,DDias08,Mello08,Caixa07}.
Here we study the EPS profile as the parameter $A$ changes
from $A=0$  near $T=T_{PS}$ to $A=1$ close to $T=0$K.

\begin{figure}[!ht]
\begin{center}
     \centerline{\includegraphics[width=7.0cm]{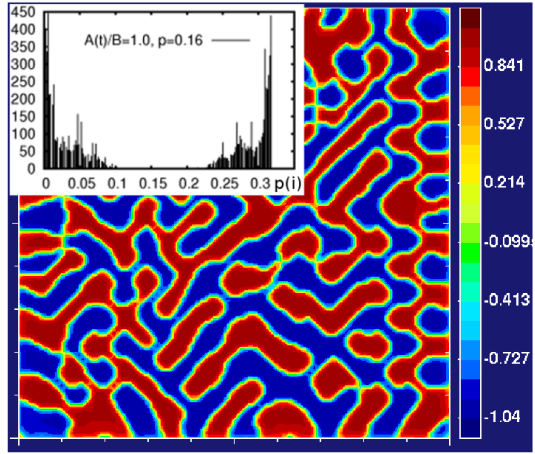}}
\caption{ (color online) The order parameter $u(i)\equiv (p(i)-p)/p$  
simulation of the inhomogeneous local charge density on a $100 \times 100$ sites. In the
inset, the local densities $p(i)$ histogram of the optimal compound ($p=0.16$) 
indicating the tendency to segregate into grains with $p(i) \approx 0$ and 
$p(i)\approx 0.32$. As the temperature $T$ goes down below the phase separation
temperature $T_{ps}$ the segregation increases.
} 
\label{MapHist} \end{center} \end{figure}

In Fig.(\ref{MapHist}) we show a typical density profile after
a long time evolution  ($A,B=1$ and average doping $p=0.16$) 
with a high segregation level as it is seen by the local 
densities histogram (in the inset).
The colour map shows regions with $p(i)\approx 0$ (red) and 
$p(i) \approx 2p$ (blue).
The lack of direct measurements does not let us know whether all cuprates 
have such high level of charge inhomogeneity and it is possible that it 
varies according to a specific family of compounds.

\section{The EPS and the Fundamental Interaction}

\begin{figure}[!ht]
\begin{center}
   \centerline{\includegraphics[width=6.0cm]{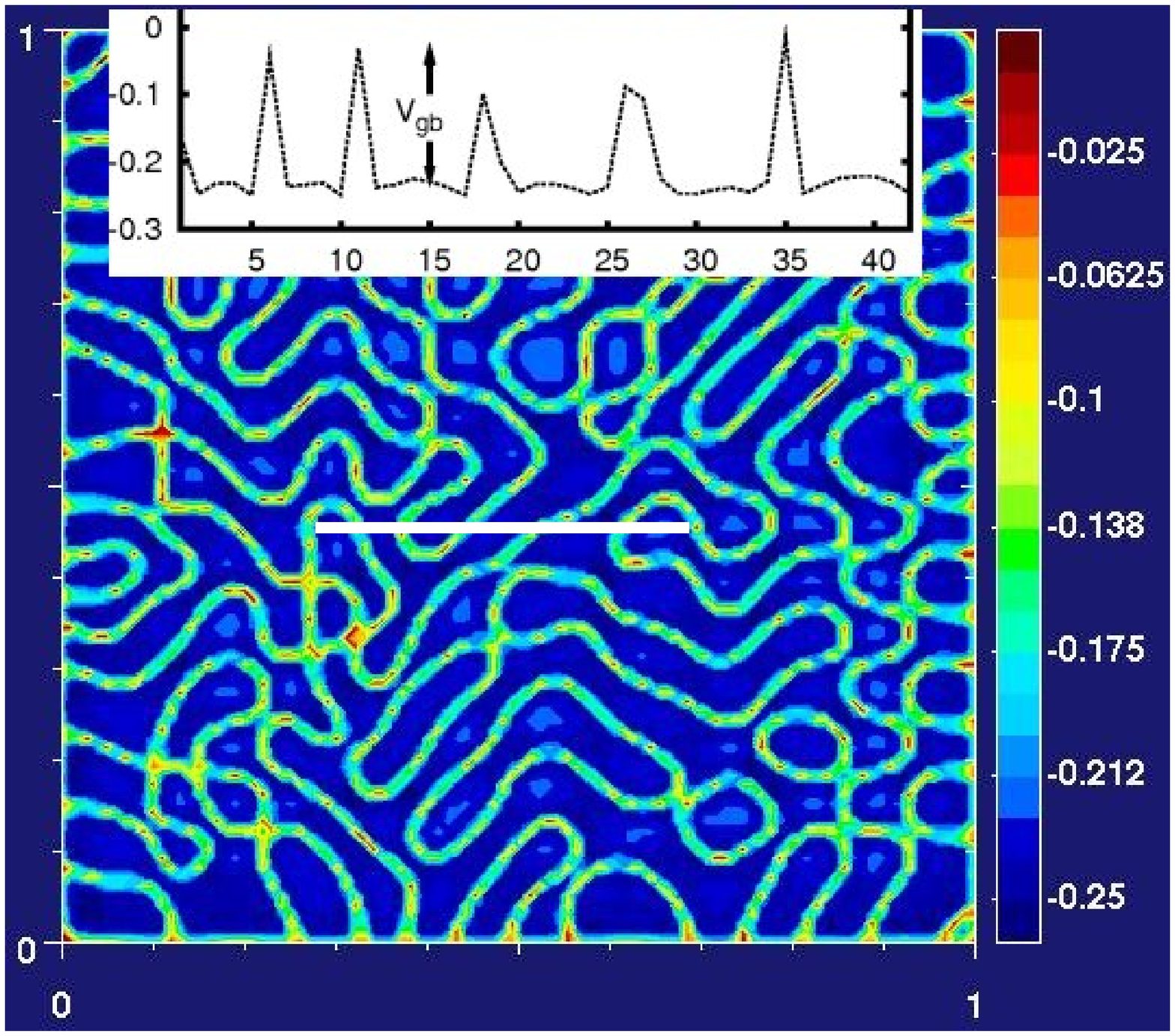}}
   \centerline{\includegraphics[width=6.0cm]{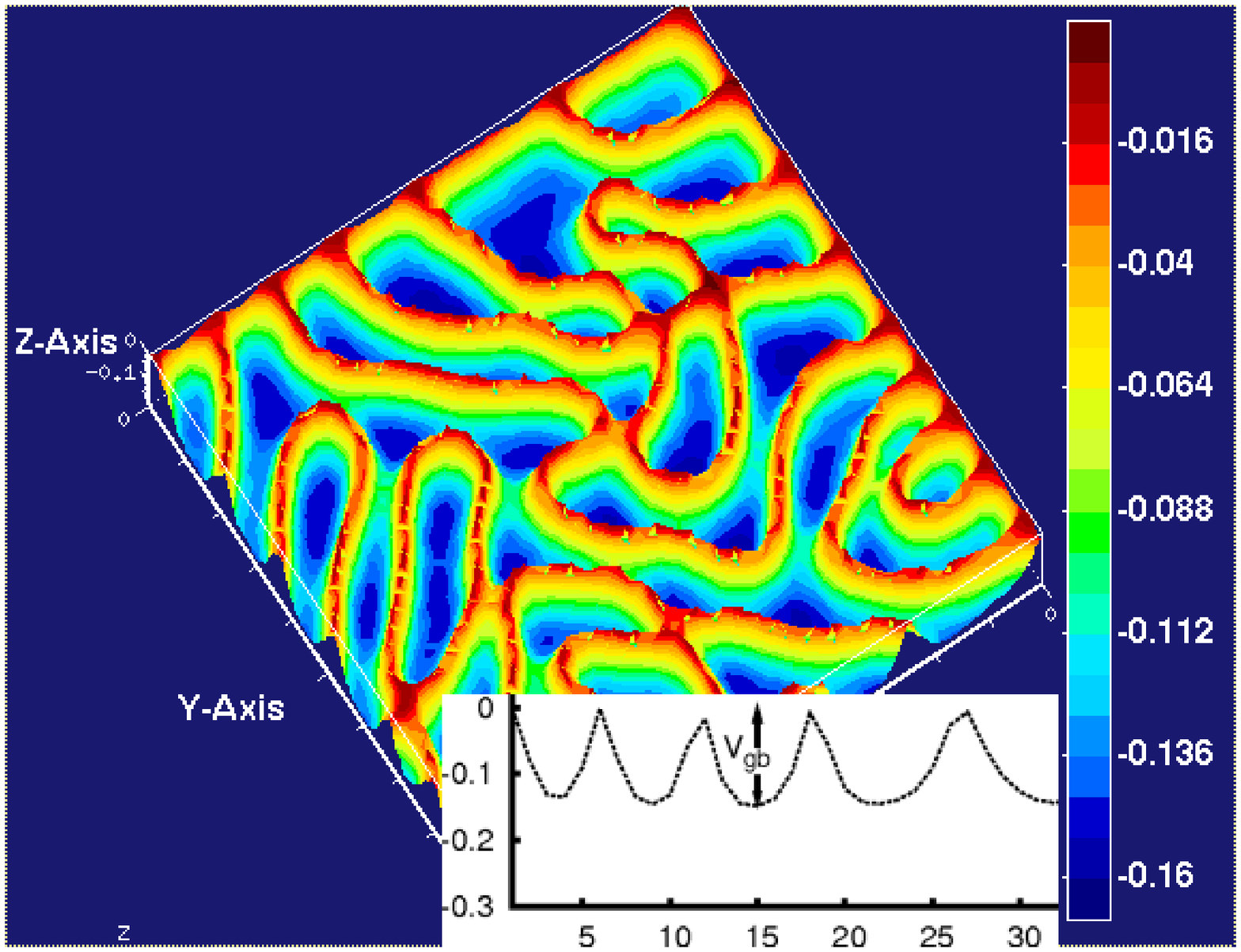}}
\caption{ (color online) On the top panel, the  potential ${\it V}(p,i,T)$ 
planar map simulation. The  values on 42 sites at the white line
are shown in the top inset to demonstrate the potential wells with
average barrier $\approx V_{gb}$ where single-particle bound states 
appear. The lower panel is a 3-dimensional section of the above map,
where the potential barriers are more clearly visible, 
at a higher temperature and  smaller $V_{gb}$, as shown in the
inset.
}
\label{EV6200}
\end{center}
\end{figure}

During the EPS process the holes move preferable along the $Cu-O$ 
bonds or  nearest neighbor hopping according to the measured 
dispersion relations\cite{Schabel}. Here we show by the 
local free energy calculations
that this hole motion can be regarded as if it was produced
by an effective hole-hole interaction as has been proposed
before, like, for instance, by the work of Trugman\cite{Trugman}. To see this we
perform numerical simulations of the local potential energy
$V_{GL}(u(i,T))$ with the solutions $u(i,T)$ of the CH equation. 
The results are shown in Fig.(\ref{EV6200}) and they 
reveal two important and different effects:
{\it i-} it divides the system in many tiny potential wells as it is shown 
on the free energy map of Fig.(\ref{EV6200}). The inset shows also the 
free energy along the white straight line on 
42 sites ($\approx 160\AA$) showing these potential wells and
the barriers between the low and high density regions. 
We define the height of these inter-grain or grain-boundary potential
$V_{gb}$ since it gives a granular structure to the system. 
In this way each small grain of low or high density is a small 
bounded region by $V_{gb}$. Each of these potential wells form 
local single-particle  bound states. These bound states are seen 
experimentally by the local density of states (LDOS) derived 
from
STM experiments\cite{McElroy,Gomes,Pasupathy,Kato,Pushp,Kato2}. 
$V_{gb}(T)$ is function of the temperature and increases when the
temperature goes down below $T_{PS}$ as it is demonstrated by the
two insets in Fig.(\ref{EV6200}) taken from simulations at 
two different temperatures. \\
{\it ii-} Second, in the process of minimizing 
the local free energy, the holes move 
to cluster themselves in a similar fashion as if they attract
themselves, forming the hole rich and hole poor regions.

\begin{figure}[!ht]
\begin{center}
    \begin{center}
     \centerline{\includegraphics[width=4.0cm]{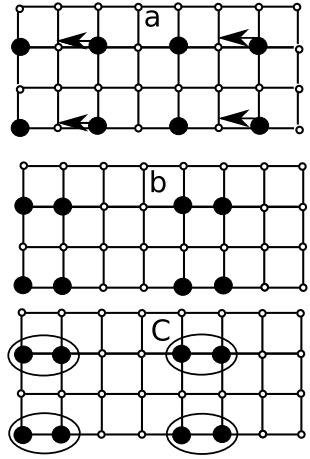}}
    \end{center}
\caption{Schematic representation of the cluster and pair
formation. a) represents a homogeneous system with $p=0.25$
(one hole at each four sites) and the CH phase separation
produces hopping preferably along the $Cu-O$ bond
toward clusters formation of low ($p_i=0$)
and high densities ($p_i=0.50$)(b). At low temperatures, d-wave
superconducting pairs may be formed (c).  }
\label{VintPair}
 \end{center}
\end{figure}

This hole-hole effective attraction is schematically illustrated in
Fig.(\ref{VintPair}) where a) represents
a homogeneous system with $p=0.25$ (one hole at each four sites)
and b) the motion toward clusters formation of low ($p_i=0$) and
high densities ($p_i=2p=0.50$). This hole movement can be regarded 
as originated from {\it an effective two-body attraction}. Conceptually, this
is similar to the spin-spin exchange interaction, that arises from the Pauli principle and
the minimization of the local electronic energy, that produces ferromagnetic
order. 

We made simulations using ${\it V_{GL}}(p,i,T)$ at low temperatures
up to $T_{PS}(p)$, where the disorder disappears. The low temperatures
values are parameterized to yield the low temperature gaps at different 
average doping $p$ from the STM measurements
of the series of Bi2212 LDOS\cite{McElroy}.  In this way we obtain a numerical 
estimate of $V_{gb}(p,T)$ that reproduces the STM results of
this entire family. Namely
\begin{eqnarray}
V_{gb}(p,T)=V(p)V(T)=(-0.7+2.4p)(1-(T/T_{PS})^{1.5}),
\label{VpT}
\end{eqnarray}
where the values are in $eV$, $V(p)$ is linear and vanishes
at $p\approx 0.32$ following the approximate behavior of
$T_{PS}(p)$ from the $T_0(p)$ curve plotted in the phase 
diagram of Timusk and Statt\cite{TS}. 
The power to the 1.5 temperature dependence is taken
from the work of Cahn and Hilliard\cite{CH}. 

\section{BdG-CH Combined Calculations}

We now perform self consistent-calculations
with the Bogoliubov-deGennes (BdG) theory in the system.
There are some possibilities to obtain a solution with 
a disordered density and local dependent superconducting gaps, for
instance: Ghosal et al\cite{Ghosal}
introduced a varying impurity potential in the chemical potential to
assure the local changes in the hole density. Other approach
is to introduce modulations to
the superconducting pair interaction due to the out of plane
dopant atoms\cite{Nunner}. Here we introduce a third method, different 
but it contains the spirit of the work of Ghosal: we keep the local
disordered density solution derived from the CH simulation 
fixed at all points and determine self-consistently the respective 
local chemical potential and d-wave superconducting amplitudes. 
In this way we capture the phase separation solution  like 
that shown in Fig.(\ref{MapHist}). To calculate the
superconducting amplitudes and single-particle excitations
and bound states we used the effective hole attraction $V_{gb}$ 
in the form of a local two-body nearest neighbor attraction. 
Although the potential is constant in space, the different local
densities yield different superconducting amplitudes.
Starting with an extended Hubbard Hamiltonian with nearest 
neighbor hopping $t=0.15$eV and next nearest
neighbor hopping of $t_1/t=-0.64$, close to the ARPES
value of $t_1/t=-0.50$\cite{Schabel}, the BdG mean-field equations
are written in terms of the BdG 
matrix\cite{Mello04,DDias07,DDias08,Mello08,Caixa07}:

These equations, defined in detail in Refs.\cite{Mello04,DDias08},
are solved self-consistently for $E_n \ge 0$ together with the
eigenvectors $(u_n,v_n)$ and d-wave pairing amplitudes
\begin{eqnarray}
\Delta_{d}({\bf x}_i)&=&-{V_{gb}\over 2}\sum_n[u_n({\bf x}_i)v_n^*({\bf x}_i+{\bf \delta})
+v_n^*({\bf x}_i)u_n({\bf x}_i  \nonumber \\
&&+{\bf \delta})]\tanh{E_n\over 2k_BT} ,
\label{DeltaV}
\end{eqnarray}
and the input inhomogeneous hole density is given by
\begin{eqnarray}
p({\bf x}_i)=1-2\sum_n[|u_n({\bf x}_i)|^2f_n+|v_n({\bf x}_i)|^2(1-f_n)],
\label{density}
\end{eqnarray}
and converges self-consistently to a $N \times N$ square (here
we made calculations with $N=28-42$)
in the CH density map of Fig.(\ref{MapHist}). $f_n$ is the Fermi function. 
$V_{gb}$ is maintained fixed at each temperature $T$ and for a given
compound with average doping $p$.

The BdG-CH combined calculations yield larger superconducting
amplitudes $\Delta_d(i,p,T)$ in the high density grains and
smaller amplitudes at regions with low densities. The variations
on $\Delta_d(i,p)$ are usually $\pm 12\%$ around
the average value $\Delta^{av}_d(T,p)\equiv \sum_i^N \Delta_d(T,i,p)/N$. 
However, due to the mean-field approach, all the amplitudes vanishes
at the same temperature $T^*(p)$. We believe that a more rigorous treatment
would yield that larger gaps vanish at larger temperatures, but there
are many interesting consequences even within this simple approach.
Due to the functional form of $V_{gb}(p)$ in Eq.(\ref{VpT}) the
$\Delta^{av}_d(T,p)$ decreases systematically with $p$ as it is shown in Fig.(\ref{EJTc}).
As the temperature decreases below $T^*(p)$  the
grains become superconductors but the grain boundary potential 
barrier $V_{gb}$ prevents the current to flow freely. Thus, the 
electronic grain structure of cuprate superconductors at low temperatures
can be regard as formed by numerous $S_i-I-S_j$ junctions.

Consequently, we apply the theory of granular superconductors\cite{AB}
to these tiny electronic grains, and as an approximation, we  use
the Josephson coupling expression to an $S-I-S$ junction 
given by\cite{AB}.
\begin{eqnarray}
E_J(p,T) = \frac{\pi h\Delta(T,p)}{4 e^2 R_n}
tanh(\frac{\Delta(T,p)}{2K_BT_c}).
\label{EJ}
\end{eqnarray}
$\Delta^{av}_d(T,p)$
is the average of the calculated BdG superconducting
gaps $\Delta_d(i,T)$ on the entire $N \times N (N=28-42)$ square.
The $R_n$ is the normal resistance of a
given compound, which we take as proportional to the $\rho_{ab}$
measurements\cite{Takagi} on the complete series  of
$La_{2-p}Sr_pCuO_2$. The values of $\Delta^{av}_d(T,p)$ as function
of $p$ are plotted in the top panel of Fig.(\ref{EJTc}). In the
low panel we plot the two sides of Eq.{\ref{EJ}) and the
intersections yield one of our main result: the
$T_c(p)$ dome shape in agreement with the Bi2212 series.

\begin{figure}[!ht]
\begin{center}
  \begin{minipage}[b]{.1\textwidth}
    \begin{center}
      \centerline{\includegraphics[width=7.0cm]{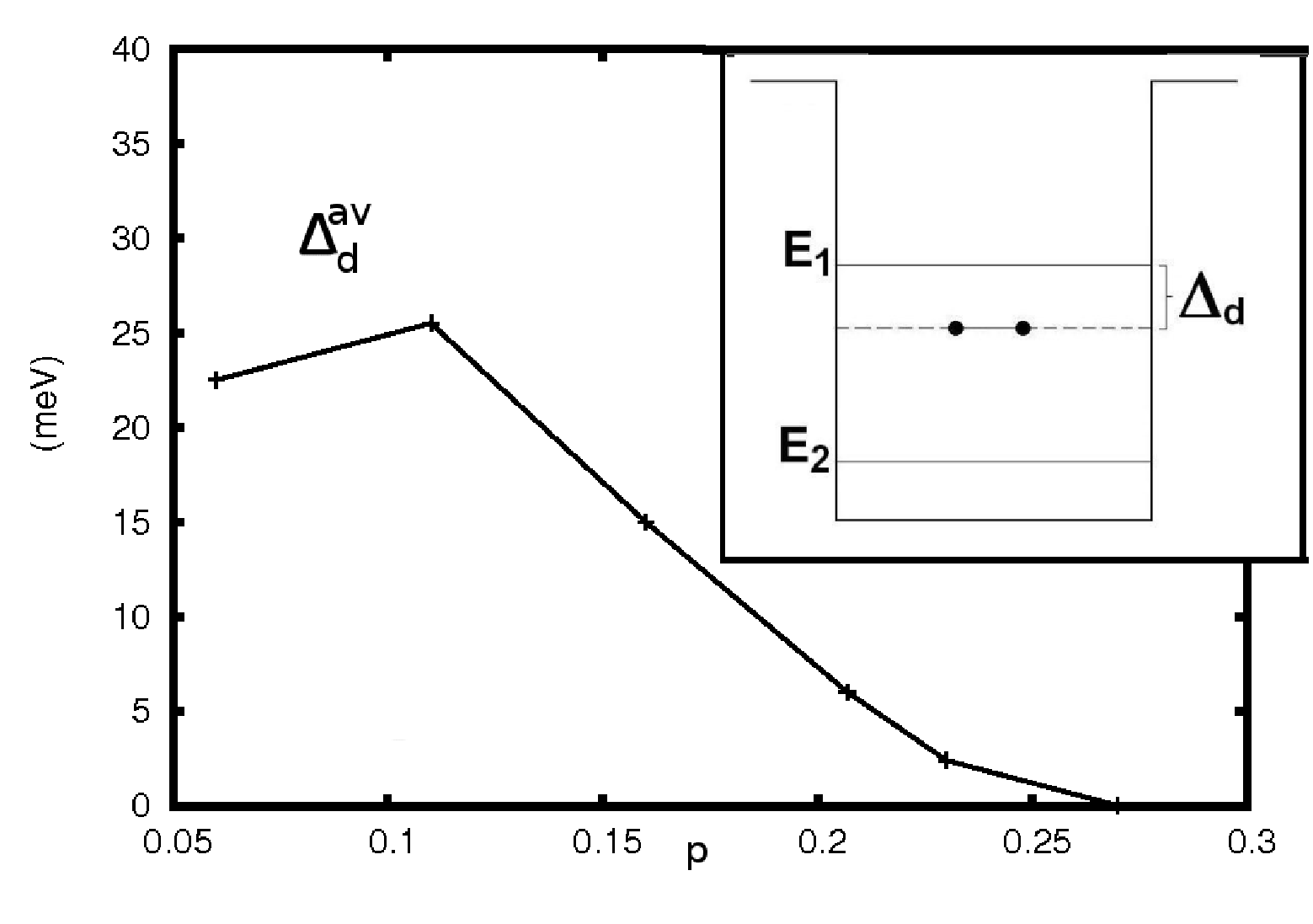}}
    \centerline{\includegraphics[width=7.0cm]{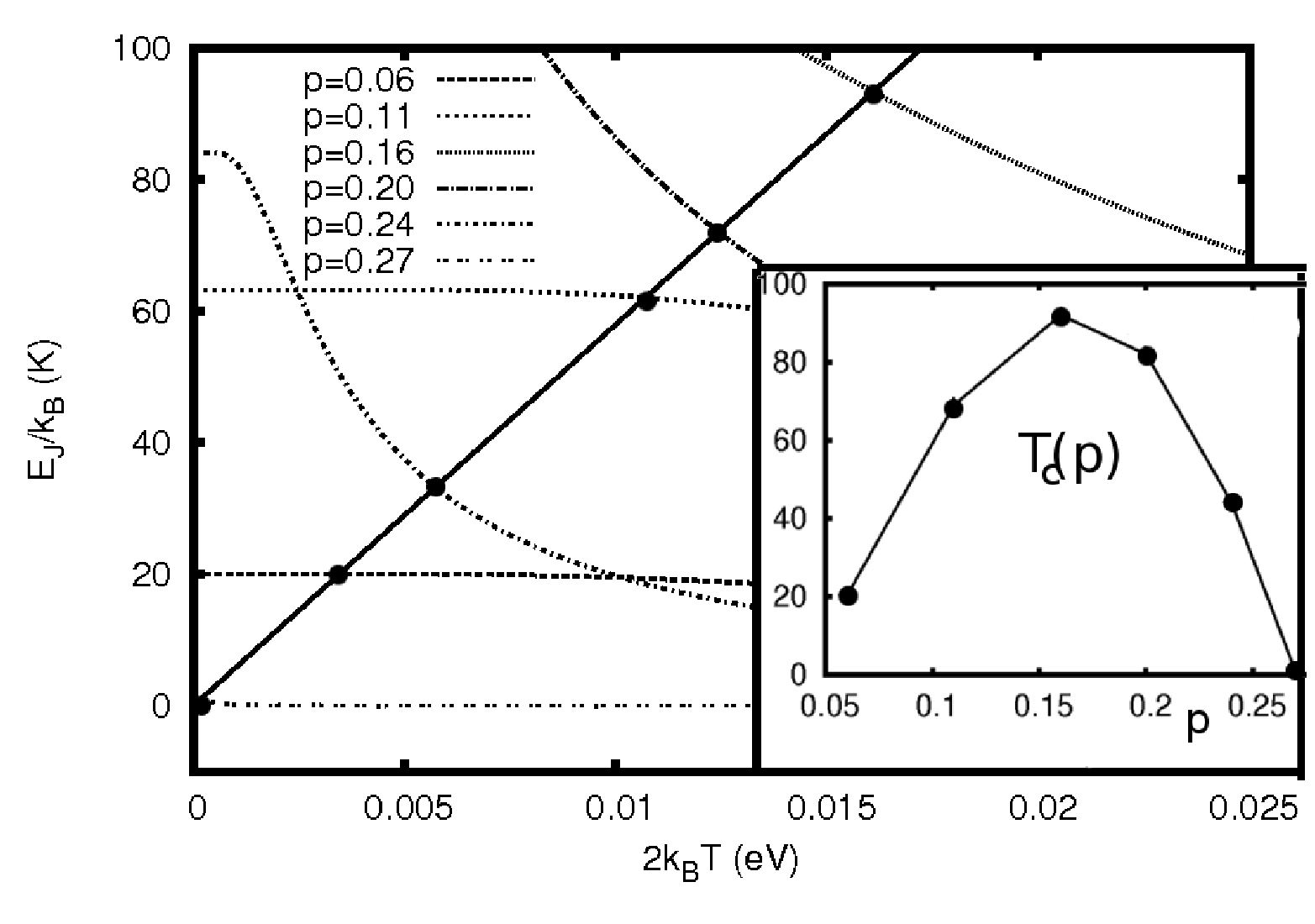}}
   \end{center}
  \end{minipage}
\caption{In the top panel we plot the average $\Delta_d(T,p)$ and
in the inset
the schematic single particle energies that appear in the LDOS. In the
low panel, the Josephson coupling among superconducting grains $E_J(p,T)$
for some selected doping values as function of T and $k_BT$. The curves intersections
give the dome shape $T_c(p)$, as plotted in the inset.  }
\label{EJTc}
 \end{center}
\end{figure}

The zero resistivity transition takes place when the Josephson
coupling $E_J$ among these tiny grains is
sufficiently large to overcome thermal fluctuations\cite{Merchant}, i.e.,
$E_J(p,T=T_c) \approx k_BT_c(p)$, what leads to phase locking and
long range phase coherence. Consequently {\it the  superconducting
transition in cuprates occurs in two steps}, similar to a
superconducting material embedded  in a non superconducting
matrix\cite{Merchant}: First, as the temperature goes down, by the 
appearing of intragrain superconductivity ({\it pseudogap phase}) and 
by Josephson coupling with phase locking ({\it superconducting phase}).
We emphasize that the reason for the dome shape form of $T_c(p)$, 
with the optimum value near $p=0.16$, is due to the fact that 
$\Delta^{av}_d(T,p)$ decreases and $(1/R_n$) increases with $p$.

\section{The STM Results and Interpretation}

One of the most important experimental result without a widely 
accepted explanation is the spatial dependent energy gaps $\Delta(\vec r)$ 
measured by Scanning Tunneling Microscopy 
(STM)\cite{McElroy,Gomes,Pasupathy,Kato,Pushp,Kato2}. We show here that this
behavior can be reproduced by the microscopic granular theory
developed above. We perform calculations on the non-uniform charge system but
the usual  local density of
states (LDOS)\cite{Gygi} at different places ${\bf x}_i$ are not so sharp and difficult
to get their real value. To improve the determination of the peaks positions, although
the measured LDOS of cuprates are not symmetric,
we deal with the following symmetric LDOS,

\begin{eqnarray}
 N_i(E)=\sum_n&[&|u_n({\bf x}_i)|^2 + |v_n({\bf x}_i)|^2]
\times  \nonumber \\
&& [f_n^{'}(E-E_n)+f_n^{'}(E+E_n)].
\label{LDOS}
\end{eqnarray}
$f_n$ is the Fermi function,
the prime is the derivative with respect to the argument,
and $u_n, v_n$ and $E_n$ are respectively the eigenvectors
and positive eigenvalues (quasi-particles
exciting energy) of the BdG matrix
equation\cite{Mello04,DDias07,DDias08,Mello08,Caixa07}. 

To study the effects of the  
disordered density, we examine the ratio $LDOS(V_{gb}\ne
0)/LDOS(V_{gb}=0)$. $LDOS(V_{gb}=0)$  contains the 
inhomogeneous charge distribution and $LDOS(V_{gb}\ne 0)$, the
unnormalized local density of states, vanishes 
around the Fermi energy due to the superconducting 
gap and the single particle levels in the grains. 
This LDOS ratio yields well-defined peaks 
and converges to the unity at large bias,
in close agreement to similar approach of the  
LDOS ratios calculated from STM 
measurements\cite{Pasupathy}. 

These features are illustrated in Fig.(\ref{LDOS36n20T40R}) 
for an overdoped $p=0.20$ compound near a grain boundary at
a representative  average doping hole point. At $T=40$K, 
$V_{gb}=0.240$eV  and for large applied potential difference 
($V>0.1$eV) both LDOS, $LDOS(V_{gb}\ne 0)$ and
$LDOS(V_{gb}=0)$ converge to the same values, but around zero bias
($V\approx 0$) there is a large spectral weight suppression
in $LDOS(V_{gb}$ only. This is due to the single-particle
bound states and to the superconducting pair formations and
both effects are destroyed as the temperature is raised,
according to the lower panel of Fig.(\ref{LDETp}).
To distinguish clearly the effect of the superconducting gap 
in the DOS of Fig.(\ref{LDOS36n20T40R}) we draw arrows at the 
$\Delta_d$  values. Notice that it appears only 
as an anomaly in the LDOS curve with two well-defined peaks, as
it was reported by some STM data\cite{Kato,Pushp}. Consequently,
the gap calculated directly from the LDOS peaks is identified with
the  with the pseudogap $\Delta_{PG}(p)$. This also implies that
STM may not measure the local superconducting gap but, in some cases,
the local gap from the single particle bound states
formed in the isolated islands.

\begin{figure}[ht]
\begin{center}
     \centerline{\includegraphics[width=6.0cm,angle=-90]{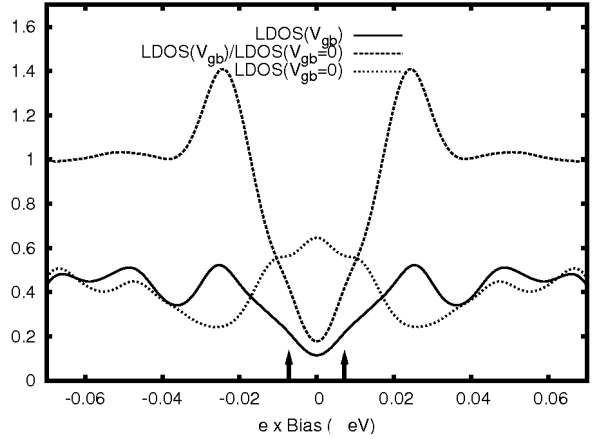}}
\caption{ The real LDOS with $V_{gb}=0.240$eV, the zero pair potential LDOS $V_{gb}=0$ 
and the normalized LDOS on an overdoped $p=0.20$ sample 
for $V_{gb}=0.240$eV. The zero pair potential LDOS $V_{gb}=0$ is also
shown. The arrows show the values of  the local superconducting gap
$\Delta_d=7.2meV$, smaller than the LDOS gap ($\approx 24$meV). 
} 
\label{LDOS36n20T40R} 
\end{center}
\end{figure}

In Fig.(\ref{LDETp}) we plot the calculated low temperature LDOS  at four
representative locations with different densities of the $p=0.16$
optimum compound. We see that the peaks vary from $35-65meV$ in
accordance with the data of McElroy et
al\cite{McElroy}. This overall agreement comes from the choice of the
parameter in the potential $V_{gb}$ (Eq.\ref{VpT}), however the local
variations in nanometer scale comes from the charge inhomogeneity
calculated from the CH approach
and also reproduces the {it local variations} of the STM LDOS\cite{McElroy}.
The temperature evolution in the low panel of Fig.(\ref{LDETp})
shows that the single-particle or LDOS gap closes near $T^*=115$K,
more than 20K above $T_c$. 

\begin{figure}[!ht]
\begin{center}
  \begin{minipage}[b]{.1\textwidth}
    \begin{center}
    \centerline{\includegraphics[width=5.0cm,angle=-90]{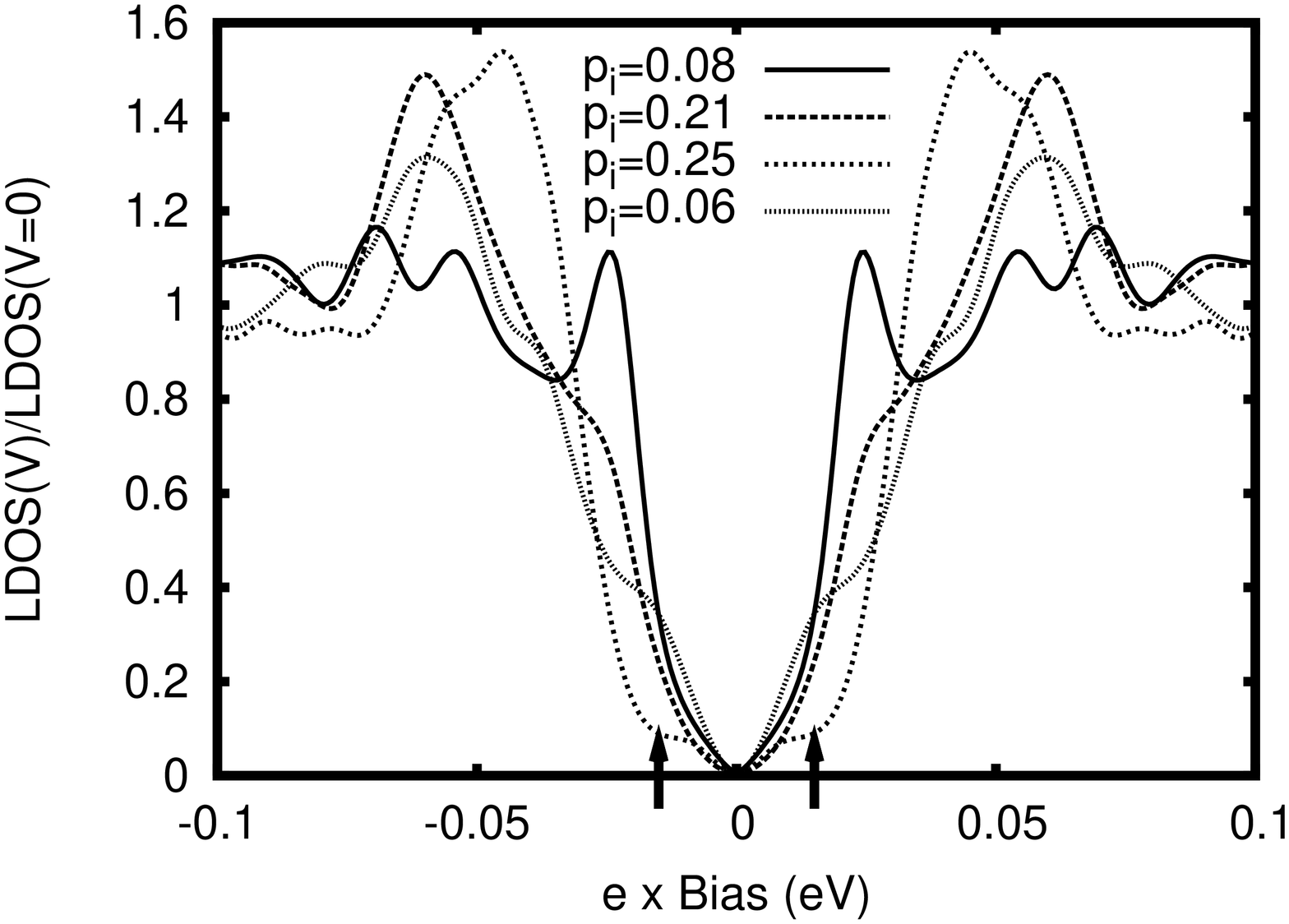}}
    \centerline{\includegraphics[width=5.0cm,angle=-90]{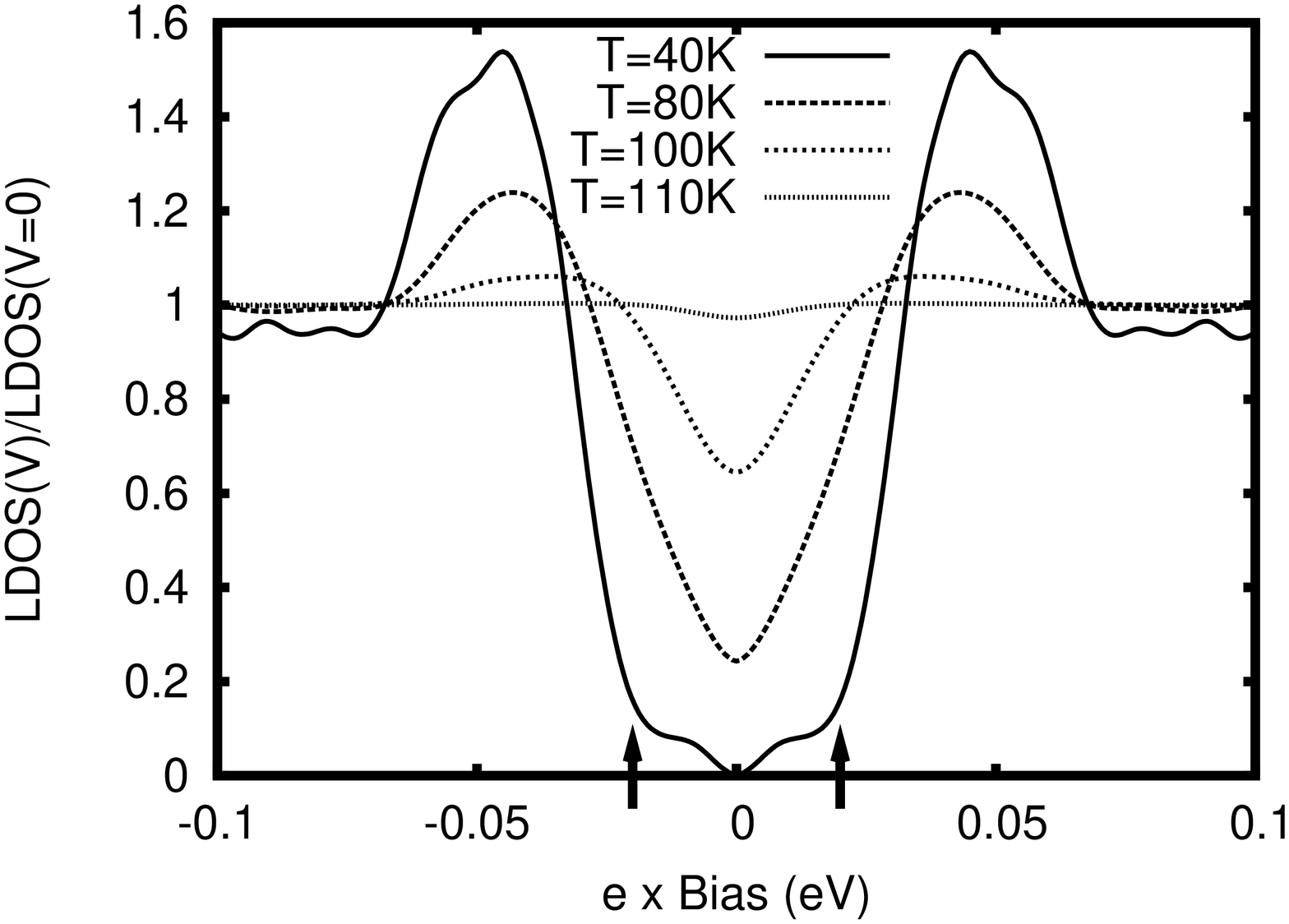}}
    \end{center}
  \end{minipage}
\caption{In the top panel, the LDOS at four different regions with
local doping $n_i$.  At the low panel the temperature evolution of
the LDOS just to the closing of the gap at $T\approx 114$K, above
$T_c\approx 92$K. The arrows show the average
$\Delta^{av}_d(T=40K)=15meV$. In low panel we show the LDOS
temperature evolution at one region ($p_i=0.25$). The kink at low
bias in the low temperature curve has also been observed by many STM
data\cite{Gomes,Pasupathy,Kato,Pushp,Kato2} and in our 
calculations it is due to the
superconducting gap $\Delta_d(T=40K)=21meV$. }
\label{LDETp}
\end{center}
\end{figure}

The calculations in the density disordered system 
yield that the low temperature superconducting
gaps $\Delta_{d}(i,p=0.16)$ vary from $7-23meV$, depending 
on the local density, that is, they are much smaller than the 
gaps derived from the LDOS peaks shown in Fig.(\ref{LDETp}). 
Consequently we conclude that
the measured LDOS gaps are due to the single-particle bound
states gaps $\Delta_{PG}(i,p,T)$ that leads to the following
scenario to the cuprate superconductors:
The pseudogap phase derived from our calculations is
formed due to the granular behavior by the single-particle bound
states gaps $\Delta_{PG}(i,p,T)$ in the isolated local doping
dependent regions or grains. Like in granular superconductors, 
there are also the intragrain superconducting
amplitude $\Delta_{d}(i,p,T)$. At lower temperatures the
superconducting phase also contains both gaps but has phase coherent
through the Josephson coupling among the grains. At the overdoped
region $T^*(p)$ approaches $T_c(p)$ while their difference increases
in the underdoped region. These two calculated curves as function of $p$
are shown in the phase diagram of Fig.(\ref{PhaseDiag}). 

We have also plotted the EPS line $T_{PS}(p)$ used in our calculations
where stats the phase separation process and  which we assumed to be near 
the observed anomaly calledupper-pseudogap Timusk and Statt\cite{TS}. 
As mentioned in the introduction, this
transition may be due to the presence of the AF order which
has lower free energy\cite{Mello09}
(intrinsic origin) or due to out of plane arrangements of 
dopant atoms\cite{Keren,Curro,Bianconi2} (extrinsic origin).

\begin{figure}[!ht]
     \centerline{\includegraphics[width=6.0cm,angle=-90]{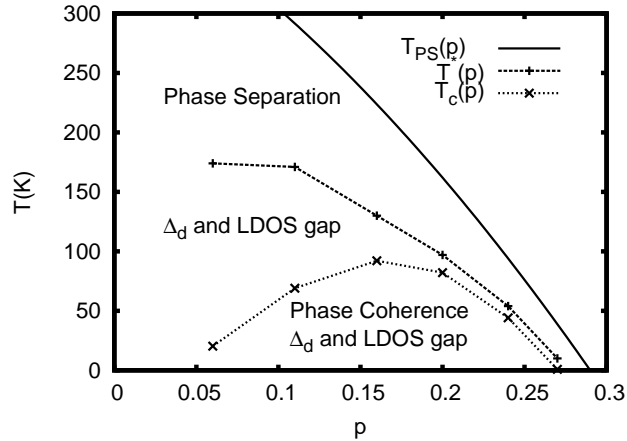}}
\caption{ The calculated phase diagram of cuprate superconductors as
derived from the EPS transition $T_{PS}(p)$ and the formation of
tiny grains. The onset of single particle bound states and
superconducting amplitude formation occurs at $T^*(p)$. The dome
shape $T_c(p)$ curve is due to Josephson coupling among the grains.
These three lines are in agreement with the experimental results. }
\label{PhaseDiag} 
\end{figure}

\section{Conclusions}

The main ideas of our work come from the application of 
the CH theory to model the highly non-uniform
charge distribution in cuprates. This approach describes
the formation of local isolated regions of low and high densities
formed at the many free energy potential minima. These non-uniform 
regions act as attraction centers and reduces the charges kinetic energy
what favours the Cooper pair formation.
We show that an effective hole-hole attraction 
$V_{gb}(p)$ appears from the Ginzburg-Landau free energy and 
the Cahn-Hilliard simulations. These simulations reveal also 
the microscopic granular structure that leads to the
charge confinement and the 
single-particle bound states (as seen in Fig.(2)). With the values
of $V_{gb}(p)$ matching the low temperature Bi2212 average LDOS gap
values of McElroy et al\cite{McElroy}
and the assumption of long range order by Josephson coupling
among the different regions we
reproduce the entire phase diagram shown in Fig.(\ref{PhaseDiag}).
The dome shape form of $T_c(p)$
is a consequence of the different behavior of $1/R_n(p)$ which
increases, and $\Delta^{av}_d(p)$ which decreases with $p$.

In summary: the LDOS main peaks are likely to be due to 
the intragrain single-particle bound states $\Delta_{PG}$ and the
local superconducting gap $\Delta_d$ produces in general only a small 
anomaly at low temperature and low bias as observed by many STM
data\cite{Gomes,Pushp,Kato,Kato2}.
$\Delta_d$  and $\Delta_{PG}$ have completely different nature but
they have the same origin, namely, the inhomogeneous potential
$V_{gb}$. They are both present in the pseudogap and superconducting
phases but they have distinct strengths. Their differences were verified in the
presence of strong applied magnetic fields and also by  their
distinct temperature behaviour in tunneling
experiments\cite{Krasnov}.
The potential barriers $V_{gb}$
prevents phase coherence and, as in a granular superconductor, the
resistivity transition occurs due to Josephson coupling  among the
intragrain superconducting regions (Eq.\ref{EJ}). 

\section{Acknowledgment}
We gratefully acknowledge partial financial aid from Brazilian
agencies CNPq and FAPERJ.
%

\end{document}